\documentclass[preprint,12pt]{elsarticle}

\usepackage{amssymb}
\usepackage{amsmath}
\usepackage{graphicx}
\usepackage{nomencl,etoolbox,ragged2e,siunitx,mathtools}
\usepackage{setspace}
\usepackage{booktabs}
\usepackage{lscape}
\usepackage{arydshln}
\usepackage[algoruled,boxed,lined]{algorithm2e}
\usepackage{algorithmic}
\usepackage{color}
\usepackage{multirow}
\usepackage{tablefootnote}
\usepackage{physics}
\usepackage[dvipsnames]{xcolor}
\journal{Journal of Computational Physics}

\begin{document}
	
	\begin{frontmatter}
		
		\title{A scalable multi-GPU method for semi-implicit fractional-step integration of incompressible Navier-Stokes equations}
		
		\author[]{Sanghyun Ha}
		\author[]{Junshin Park and Donghyun You\footnote{Corresponding author. E-mail: dhyou@postech.ac.kr; Phone: +82-54-279-2191; Fax: +82-54-279-3199}}
		\address{Department of Mechanical Engineering, Pohang University of Science and Technology, 77 Cheongam-ro, Nam-gu, Pohang, Gyeongbuk 37673, Republic of Korea}
		
		\begin{abstract}
			A new flow solver scalable on multiple Graphics Processing Units (GPUs) for direct numerical simulation of wall-bounded incompressible flow is presented. This solver utilizes a previously reported work \cite{ha2018gpu} which proposes a semi-implicit fractional-step method on a single GPU. Extension of this work to accommodate multiple GPUs becomes inefficient when global transpose is used in the Alternating Direction Implicit (ADI) and Fourier-transform-based direct methods. A new strategy for designing an efficient multi-GPU solver is described to completely remove global transpose and achieve high scalability. Parallel Diagonal Dominant (PDD) and Parallel Partition (PPT) methods are implemented for GPUs to obtain good scaling and preserve accuracy. An overall efficiency of 0.89 is shown. Turbulent flat-plate boundary layer is simulated on 607M grid points using 4 Tesla P100 GPUs.
		\end{abstract}

		\begin{keyword}
			GPU \sep
			Boundary layer \sep
			Direct numerical simulation \sep
			Semi-implicit fractional-step method \sep
			Parallel Diagonal Dominant method
		\end{keyword}
		
	\end{frontmatter}
	
	
	\section{Introduction}
Turbulent and transitional boundary layers are comprised of a variety of scales. Such broadband scales can be captured by direct numerical simulation (DNS) which provides high-resolution data. A major challenge in DNS of such wall-bounded flows is the heavy requirement in domain length and grid size which necessitates a significant amount of computational resources. Therefore the choice of an efficient numerical scheme and algorithms for its parallelization is critical in the study of boundary layers using DNS.

A commonly used method for spanwisely periodic wall-bounded flows is the semi-implicit fractional-step method with a second-order spatial discretization. Among many variations of this method, a classic version solves the momentum equation using Alternating Direction Implicit (ADI) method, followed by the Poisson equation which is solved directly using Fourier-transform \cite{kimmoin1985}. Since there are no iterations involved, this is one of the most efficient methods for solving incompressible flow. However in parallel computing, these algorithms do not easily scale on multiple processors due to their inherently serial nature.

In a recent work \cite{ha2018gpu}, we have proposed a parallel implementation of the semi-implicit fractional-step method for Graphics Processing Units (GPUs), which represent hardwares based on a massively parallel architecture. Major difficulties coming from the serial nature of the fractional-step method have been analyzed and overcome, achieving up to $48\times$ speedup on 134M grid cells using a single Tesla P100 GPU. Yet, this work was limited to a single GPU which could only afford low to moderate Reynolds numbers.

To use this method at high Reynolds numbers, the use of multiple GPUs is inevitable. Unlike single-GPU programming which focuses on fine-grained parallelism, multi-GPU programming focuses on domain decomposition and data distribution at a coarser-level. When distributing jobs to multiple GPUs, we are interested in selecting a domain decomposition method that not only minimizes communication between GPUs but also allows coalesced access of the global memory. Note that the ADI and Fourier-transform based direct methods make frequent use of matrix transpose to efficiently access data in each direction. The problem is that matrix transpose becomes a global all-to-all operation when data is distributed on multiple GPUs. Many studies such as \cite{abide2018higher}, \cite{borrell2013}, \cite{lee2013petascale} have used global transpose in simulations of wall-bounded flows. They have obtained a weakly linear scaling on thousands of CPU cores, but reported that global transpose takes up a majority of the total computation time. Similar characteristics are reported in a recently developed GPU code \cite{zhu2018afid} whose performance depends mainly on the global transpose of the pressure solver. It will be shown later that the cost of global transpose becomes even worse when applied to the present semi-implicit fractional-step method.

The present study aims to extend the classic fractional-step method to multiple GPUs. The goal is to completely remove global transpose from the GPU algorithm, and achieve high scalability. To do so, the present study employs divide-and-conquer algorithms called Parallel Diagonal Dominant (PDD) and Parallel Partition (PPT) methods which are newly implemented to suit for GPUs. The paper is organized as follows: in Section 2, numerical methods used to discretize the governing equations are described. In Section 3, strategies for GPU implementation are explained. In Section 4, results from numerical experiments of the flow solver are provided with performance analyses. Concluding remarks follow in Section 5.

	\section{Numerical methods}
	Numerical methods are identical to those used in the previous work \cite{ha2018gpu}. Here, we offer a brief explanation which is most relevant to the present study.
	
	\subsection{Governing equations}
\begin{figure}
		\centering
		\includegraphics[width=140mm]{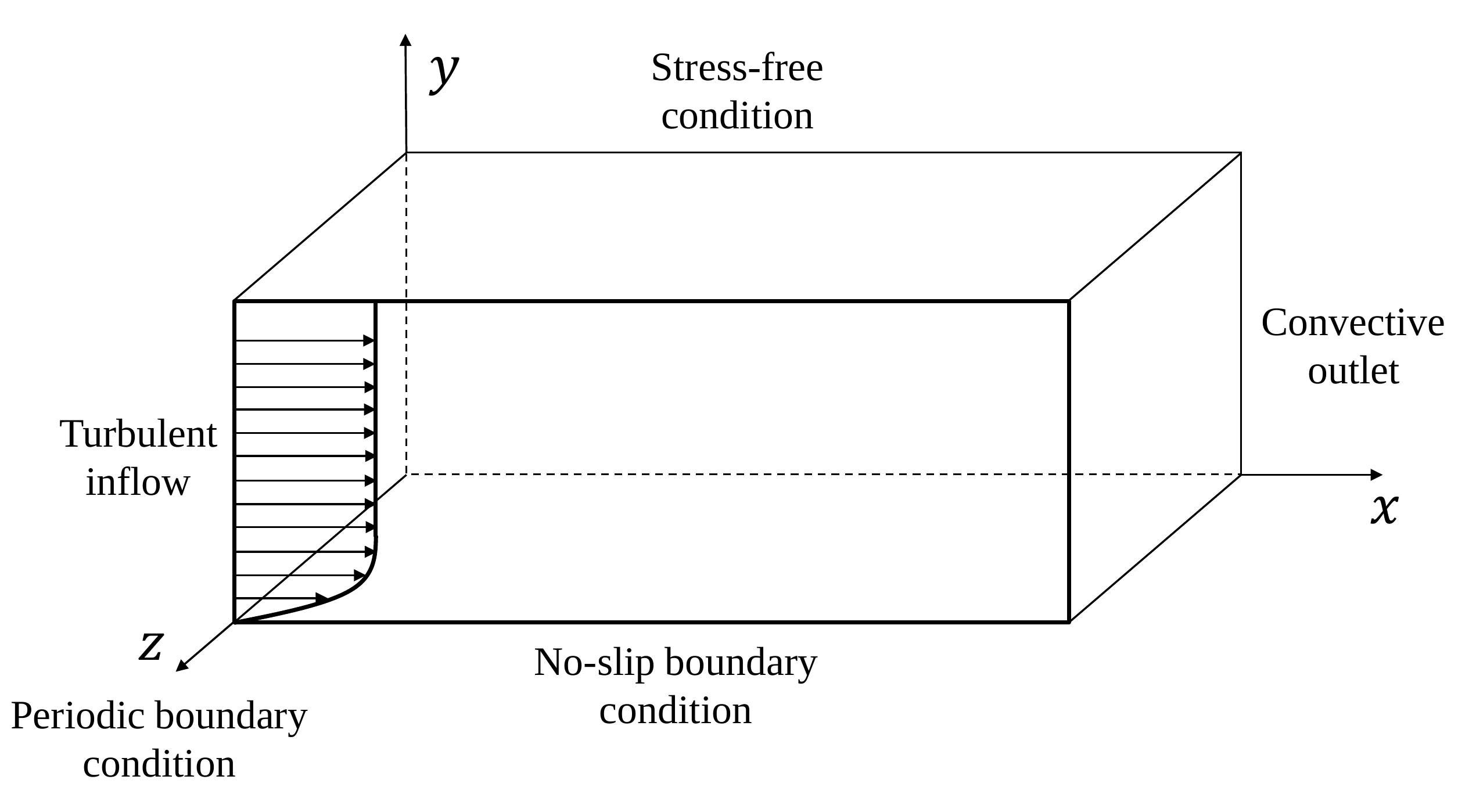}
		\caption{Flow configuration of a flat-plate boundary layer \cite{ha2018gpu}.}
		\label{fig:flowconf}
	\end{figure}

	The non-dimensionalized incompressible Navier-Stokes equations are written as
		\begin{equation} \label{eq:NScontinuity}
		\frac{\partial u_i}{\partial x_i}=0,
		\end{equation}
		\begin{equation} \label{eq:NSmomentum}
		\frac{\partial u_i}{\partial t} + \frac{\partial}{\partial x_j} u_i u_j = - \frac{\partial p}{\partial x_i} + \frac{1}{Re} \frac{\partial}{\partial x_j} \frac{\partial u_i}{\partial x_j},
		\end{equation}
	where $Re$ is the Reynolds number based on a characteristic length scale. Non-dimensional variables $u_i$ and $p$ represent velocity in the $i$-direction and pressure, respectively. Three-dimensional staggered structured grid topology is used in which all velocity components are stored at cell faces, and the pressure values at the center of each cell. Simulation of flow over a flat plate is modeled on a rectangular box (Fig.~\ref{fig:flowconf}). Uniform grid spacings are employed in the streamwise $x$- and spanwise $z$-directions, respectively, while the grid is clustered near the wall in the wall-normal $y$-direction. A no-slip condition is imposed at the bottom wall at $y=0$, and a stress-free condition at the top boundary. Convective boundary condition is applied at the outlet, while turbulent inflow is created using a recycling method~\cite{lund1998}.

	\subsection{Discretization}
	The above equations are solved by a semi-implicit fractional-step method in which the convection terms of the momentum equation are integrated explicitly in time using a low-storage third-order Runge-Kutta scheme, while the viscous terms are integrated implicitly using Crank-Nicolson scheme~\cite{hahn2002direct}. Spatial discretization is performed using second-order central difference. The momentum equation is approximated using ADI method, which produces three tridiagonal matrices for each velocity component. The Poisson equation is solved directly using half-range cosine transform in the streamwise $x$-direction and Fourier transform in the spanwise $z$-direction. Complex-numbered tridiagonal matrices in the wall-normal $y$-direction are inverted, after which the pseudo-pressure $\phi$ is obtained via inverse transforms.

	\section{GPU implementation}
	In the present fractional-step method, equations (both the momentum and Poisson) are solved in one direction at a time. This means that data orientation should also be changed whenever there is a change in the direction. Transpose operations play a critical role in communication between GPUs and the coalesced access of the global memory within each GPU. In this section, the cost of global transpose is first investigated in relation to the overall performance. Then, a different domain decomposition suitable for the present method is proposed. The present study uses Message Passing Interface (MPI) such that a 1-1 mapping between a GPU and an MPI rank is established.

	\subsection{Domain decomposition using global transpose}
\begin{landscape}
		\begin{figure}
			\centering
			\includegraphics[width=190mm]{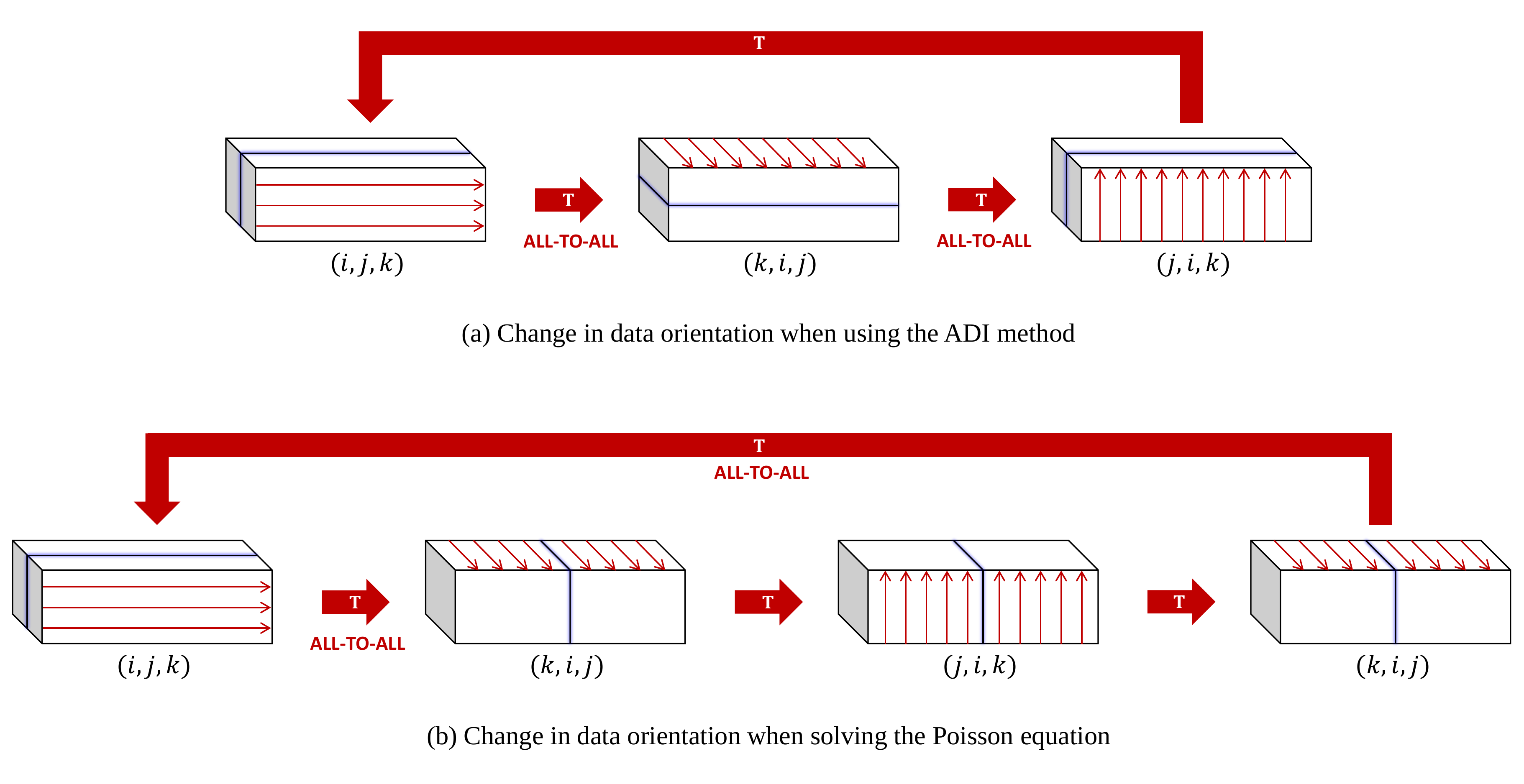}
			\caption{Local and global transposes used to change data orientation.}
			\label{fig:transpose}
		\end{figure}
	\end{landscape}

	For the purpose of testing global transpose, a one-dimensional domain decomposition in the spanwise direction is implemented. Matrices are transposed three times in the ADI method and four times in solving the Poisson equation (Fig.~\ref{fig:transpose}) as listed below. Among these, four need to be transposed in an all-to-all manner which are marked as 'ALL-TO-ALL' in Fig.~\ref{fig:transpose} and 'global' in parentheses below. Detailed implementation of local/global transpose on GPUs is based on the algorithms in \cite{CUDAfortran}.
	\begin{itemize}
		\item Momentum equation
		\begin{enumerate}
			\item Transpose x-orientation to z-orientation for z-directional ADI (global)
			\item Transpose z-orientation to y-orientation for y-directional ADI (global)
			\item Transpose y-orientation to x-orientation for x-directional ADI
		\end{enumerate}
		\item Poisson equation
		\begin{enumerate}
			\item Transpose x-orientation to z-orientation for complex-to-complex Fourier transform (global)
			\item Transpose z-orientation to y-orientation for inversion of complex-numbered tridiagonal matrices
			\item Transpose y-orientation to z-orientation for complex-to-real inverse Fourier transform
			\item Transpose z-orientation to x-orientation for complex-to-complex inverse half-range cosine transform (global)
		\end{enumerate}
	\end{itemize}

\begin{figure}
		\centering
		\includegraphics[width=140mm]{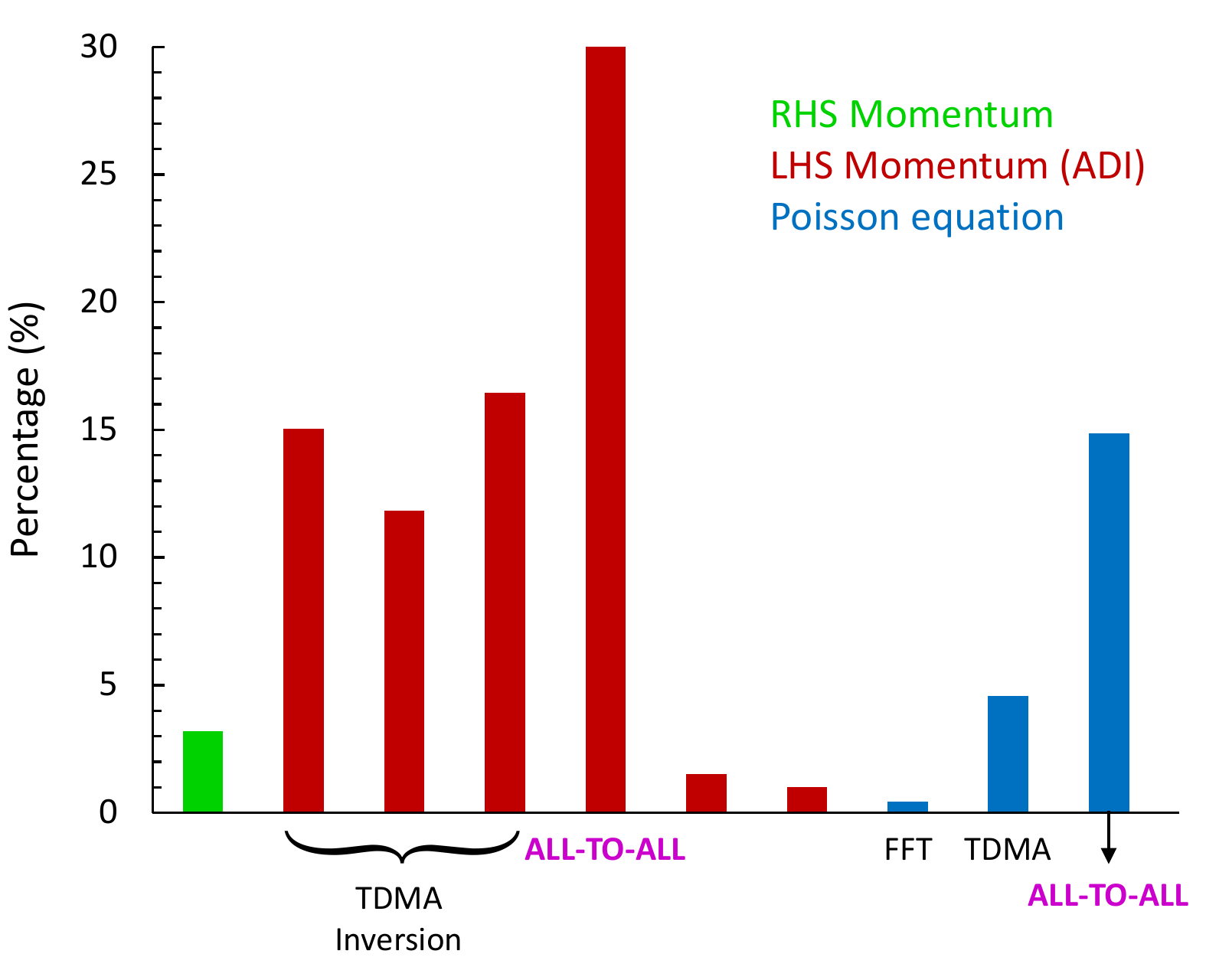}
		\caption{Relative cost of global transpose using four GPUs when compared to other parts of the flow solver. Computation time has been measured on 675M grid points.}
		\label{fig:transposecost}
	\end{figure}

	The cost of global transpose is shown in Fig.~\ref{fig:transposecost} along with other major parts of the momentum and Poisson equations. Computation time is measured on 675M grid points using 4 Tesla P100 GPUs. In both the momentum and Poisson equations, time taken to perform all-to-all communication far exceeds the main computation time such as tridiagonal matrix (TDMA) inversion or fast Fourier transform (FFT). As a result, global transpose takes up about 46\% of the entire computation time at each time-step, which makes it impractical.
	
	It should also be noted that all-to-all communication is more expensive in the ADI method compared to that of the Poisson equation. In many studies using semi-implicit fractional-step methods, only the wall-normal diffusion term of the momentum equation is integrated implicitly. Then by orienting the decomposed sub-domains in the wall-normal direction, tridiagonal matrices can be solved without all-to-all communication. This ensures that global transpose occurs only for FFT in the Poisson equation. Although this method has less communication overhead, global transpose is still the main source of reduced scalability.
		
	\subsection{Domain decomposition using parallel algorithms}
\begin{figure} 
		\centering
		\includegraphics[width=140mm]{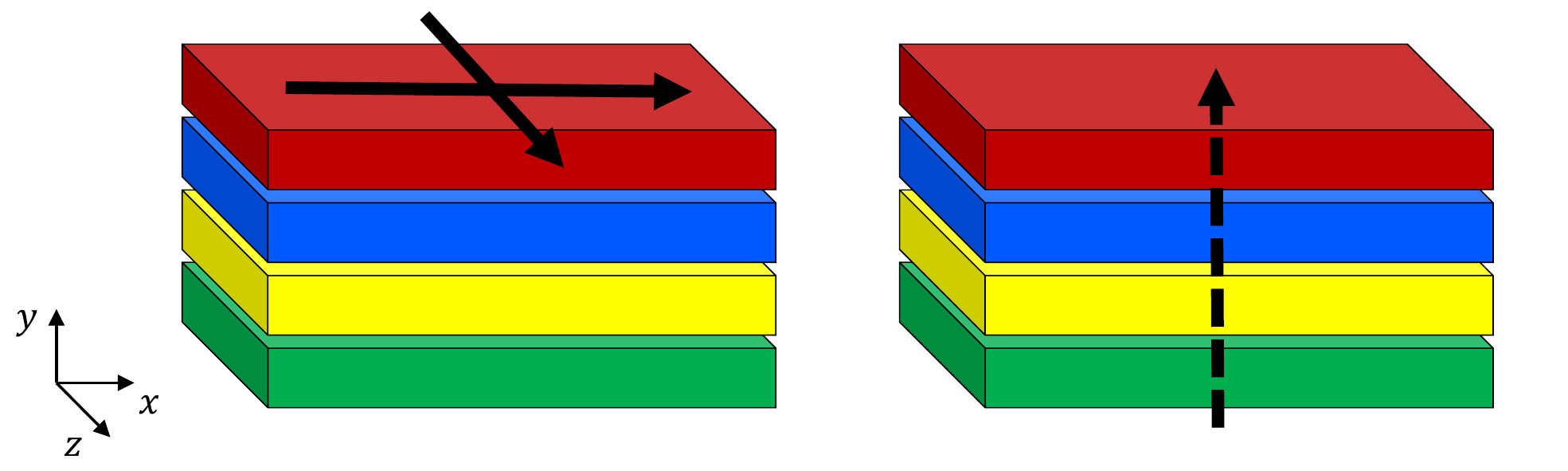}
		\caption{One-dimensional domain decomposition in the wall-normal $y$-direction. Each colored block designates a GPU. Computation in the $x$ \& $z$ directions can be carried out locally on each GPU as illustrated in the left figure. However data in the $y$-direction are scattered across different GPUs as shown in the right figure.}
		\label{fig:yslice}
	\end{figure}

	Consider a one-dimensional domain decomposition in the wall-normal $y$-direction. In this type of decomposition, FFT in $x$- and $z$-directions can be computed without global transpose, since all data resides locally in each GPU (Fig.~\ref{fig:yslice}). However data required for $y$-directional TDMAs is now distributed across different GPUs. Rather than using global transpose to collect them, two methods are employed to directly solve tridiagonal systems in parallel: the Parallel Diagonal Dominant (PDD) and Parallel Partition (PPT) methods.
	 
	PDD and PPT methods have been first proposed by Sun \textit{et al.} \cite{sun1989parallel} to solve TDMAs distributed across multiple processors. It is suited for coarse-grained parallel machines for which the number of processors is usually less than the dimension of the matrix $n$. Here, the basic idea of the algorithm is described with a specific example where $n=12$ and the number of GPUs $p=4$. For a more general and detailed derivation of this method, refer to \cite{sun1989parallel}.
	
	A tridiagonal matrix $A=[a_j, b_j, c_j]$ $(j=1,2,\cdots,n)$ can be decomposed into a block-tridiagonal matrix $\tilde{A}$ and remaining corner elements $\Delta A$.
	\begin{center}
	$A=\tilde{A}+\Delta A$.
	\end{center}
	For this example, $\Delta A$ is written as
	{\scriptsize
	\begin{center}
	$
	\Delta A = \left( 
	{\begin{array}{ccc;{2pt/2pt}ccc;{2pt/2pt}ccc;{2pt/2pt}ccc}
		\cdot&\cdot&\cdot&\cdot&\cdot&\cdot&\cdot&\cdot&\cdot&\cdot&\cdot&\cdot\\
		\cdot&\cdot&\cdot&\cdot&\cdot&\cdot&\cdot&\cdot&\cdot&\cdot&\cdot&\cdot\\
		\cdot&\cdot&\cdot&c_3  &\cdot&\cdot&\cdot&\cdot&\cdot&\cdot&\cdot&\cdot\\ \hdashline[2pt/2pt]
		\cdot&\cdot&a_4  &\cdot&\cdot&\cdot&\cdot&\cdot&\cdot&\cdot&\cdot&\cdot\\
		\cdot&\cdot&\cdot&\cdot&\cdot&\cdot&\cdot&\cdot&\cdot&\cdot&\cdot&\cdot\\
		\cdot&\cdot&\cdot&\cdot&\cdot&\cdot&c_6  &\cdot&\cdot&\cdot&\cdot&\cdot\\ \hdashline[2pt/2pt]
		\cdot&\cdot&\cdot&\cdot&\cdot&a_7  &\cdot&\cdot&\cdot&\cdot&\cdot&\cdot\\
		\cdot&\cdot&\cdot&\cdot&\cdot&\cdot&\cdot&\cdot&\cdot&\cdot&\cdot&\cdot\\
		\cdot&\cdot&\cdot&\cdot&\cdot&\cdot&\cdot&\cdot&\cdot&c_9  &\cdot&\cdot\\
		\hdashline[2pt/2pt]
		\cdot&\cdot&\cdot&\cdot&\cdot&\cdot&\cdot&\cdot&a_{10}&\cdot&\cdot&\cdot\\
		\cdot&\cdot&\cdot&\cdot&\cdot&\cdot&\cdot&\cdot&\cdot&\cdot&\cdot&\cdot\\
		\cdot&\cdot&\cdot&\cdot&\cdot&\cdot&\cdot&\cdot&\cdot&\cdot&\cdot&\cdot\\
	\end{array}}
	\right)$.
	\end{center}
	} By re-writing $\Delta A$ as $\Delta A = VE^T$, the original matrix can be written as
	\begin{center}
	$
	A = \tilde{A} + V E^T,
	$
	\end{center}
	where
	{\scriptsize
	\begin{center}
	$
	VE^T= \left( 
	{\begin{array}{cccccc}
		\cdot&\cdot&\cdot&\cdot&\cdot&\cdot\\
		\cdot&\cdot&\cdot&\cdot&\cdot&\cdot\\
		\cdot&c_3  &\cdot&\cdot&\cdot&\cdot\\
		\hdashline[2pt/2pt]
		a_4  &\cdot&\cdot&\cdot&\cdot&\cdot\\
		\cdot&\cdot&\cdot&\cdot&\cdot&\cdot\\
		\cdot&\cdot&\cdot&c_6  &\cdot&\cdot\\
		\hdashline[2pt/2pt]
		\cdot&\cdot&a_7  &\cdot&\cdot&\cdot\\
		\cdot&\cdot&\cdot&\cdot&\cdot&\cdot\\
		\cdot&\cdot&\cdot&\cdot&\cdot&c_9  \\
		\hdashline[2pt/2pt]
		\cdot&\cdot&\cdot&\cdot&a_{10}&\cdot\\
		\cdot&\cdot&\cdot&\cdot&\cdot&\cdot\\
		\cdot&\cdot&\cdot&\cdot&\cdot&\cdot\\
		\end{array}}
	\right)
	\left(
	{\begin{array}{cccccccccccc}
		\cdot&\cdot&1    &\cdot&\cdot&\cdot&\cdot&\cdot&\cdot&\cdot&\cdot&\cdot\\
		\cdot&\cdot&\cdot&1    &\cdot&\cdot&\cdot&\cdot&\cdot&\cdot&\cdot&\cdot\\
		\cdot&\cdot&\cdot&\cdot&\cdot&1    &\cdot&\cdot&\cdot&\cdot&\cdot&\cdot\\
		\cdot&\cdot&\cdot&\cdot&\cdot&\cdot&1    &\cdot&\cdot&\cdot&\cdot&\cdot\\
		\cdot&\cdot&\cdot&\cdot&\cdot&\cdot&\cdot&\cdot&1    &\cdot&\cdot&\cdot\\
		\cdot&\cdot&\cdot&\cdot&\cdot&\cdot&\cdot&\cdot&\cdot&1    &\cdot&\cdot\\
		\end{array}}
	\right)
	$
	\end{center}
	}
	We are interested in finding the solution $x$ of the system
	\begin{center}
		$Ax=d$.
	\end{center}
	This can be computed by finding the inverse of $A=\tilde{A}+VE^T$, which is given by the Sherman-Morrison matrix identity in Eq.~(\ref{eq:smidentity}).
	\begin{equation}\label{eq:smidentity}
		\left( \tilde{A} + VE^T \right)^{-1} = \tilde{A}^{-1} - \tilde{A}^{-1}V \left(  I + E^T \tilde{A}^{-1} V  \right)^{-1} E^T \tilde{A}^{-1},
	\end{equation}
	\begin{equation}\label{eq:shermanmorrison}
		x = A^{-1}d = \tilde{A}^{-1}d - \tilde{A}^{-1}V \left(  I + E^T \tilde{A}^{-1} V  \right)^{-1} E^T \tilde{A}^{-1}d.
	\end{equation}
	Since $\tilde{A}$ is block-tridiagonal, each block can be stored in each GPU. Thus $\tilde{A}^{-1}d$ and $\tilde{A}^{-1}V$ can be computed by solving the following equations locally on independent GPUs:
	\begin{equation}\label{eq:blockAx}
	\tilde{A}\tilde{x} = d,
	\end{equation}
	\begin{equation}\label{eq:blockAY}
	\tilde{A}Y = V,
	\end{equation}
	where $Y$ is written in the form of
	{\scriptsize
	\begin{center}
		$Y=
		\left( 
		{\begin{array}{cccccc}
			\cdot&w_1^{(0)}&\cdot&\cdot&\cdot&\cdot\\
			\cdot&w_2^{(0)}&\cdot&\cdot&\cdot&\cdot\\
			\cdot&w_m^{(0)}&\cdot&\cdot&\cdot&\cdot\\
			\hdashline[2pt/2pt]
			v_1^{(1)}&\cdot&\cdot&w_1^{(1)}&\cdot&\cdot\\
			v_2^{(1)}&\cdot&\cdot&w_2^{(1)}&\cdot&\cdot\\
			v_m^{(1)}&\cdot&\cdot&w_m^{(1)}&\cdot&\cdot\\
			\hdashline[2pt/2pt]
			\cdot&\cdot&v_1^{(2)}&\cdot&\cdot&w_1^{(2)}\\
			\cdot&\cdot&v_2^{(2)}&\cdot&\cdot&w_2^{(2)}\\
			\cdot&\cdot&v_m^{(2)}&\cdot&\cdot&w_m^{(2)}\\
			\hdashline[2pt/2pt]
			\cdot&\cdot&\cdot&\cdot&v_1^{(3)}&\cdot\\
			\cdot&\cdot&\cdot&\cdot&v_2^{(3)}&\cdot\\
			\cdot&\cdot&\cdot&\cdot&v_m^{(3)}&\cdot\\
			\end{array}}
		\right)
		$
	\end{center}
	}
	Here, $m=n/p=3$. The superscript denotes the MPI rank or the GPU index ranging from $0$ to $p-1$. From Eq.~(\ref{eq:shermanmorrison}), let $Z = I + E^T \tilde{A}^{-1}V$ which is a five-banded $2(p-1)\times 2(p-1)$ matrix of the form
	\begin{center}
	$Z = \left( 
	{\begin{array}{cccccc}
		1    &w_m^{(0)}&0&\cdot&\cdot&\cdot\\
		v_1^{(1)}&1    &0&w_1^{(1)}&\cdot&\cdot\\
		\hdashline[2pt/2pt]
		v_m^{(1)}&0&1    &w_m^{(1)}&0&\cdot\\
		\cdot&0&v_1^{(2)}&1    &0&w_1^{(2)}\\
		\hdashline[2pt/2pt]
		\cdot&\cdot&v_m^{(2)}&0&1    &w_m^{(2)}\\
		\cdot&\cdot&\cdot&0&v_1^{(3)}&1    \\
		\end{array}}
	\right)$.
	\end{center}
	By solving the following system for some $y$,
	\begin{equation}\label{eq:Zpdd}
		Zy=E^T\tilde{x},
	\end{equation}
	we finally obtain the solution $x$
	\begin{equation}\label{eq:xpdd}
		x = \tilde{x} - Yy.
	\end{equation}

	If we instead use a permutation matrix $P=P^{-1}$ of the form
	\begin{center}
		$P=
		\left(
		{\begin{array}{cccccc}
			\cdot&1    &\cdot&\cdot&\cdot&\cdot\\
			1    &\cdot&\cdot&\cdot&\cdot&\cdot\\
			\cdot&\cdot&\cdot&1    &\cdot&\cdot\\
			\cdot&\cdot&1    &\cdot&\cdot&\cdot\\
			\cdot&\cdot&\cdot&\cdot&\cdot&1    \\
			\cdot&\cdot&\cdot&\cdot&1    &\cdot\\
			\end{array}}
		\right),
		$
	\end{center}
	then Eq.~(\ref{eq:shermanmorrison}) becomes 
	\begin{center}
	$x = \tilde{A}^{-1}d - \tilde{A}^{-1}VP \left( P + E^T \tilde{A}^{-1} VP \right)^{-1} E^T \tilde{A}^{-1}d$.
	\end{center}
	Note that such a permutation has produced a tridiagonal $2(p-1)\times 2(p-1)$ matrix $\overline{Z}$ of the form
	\begin{center}
		$
		\overline{Z}=P + E^T \tilde{A}^{-1} VP = 
		\left(
		{\begin{array}{cccccc}
			w_m^{(0)}&1        &\cdot&\cdot&\cdot&\cdot\\
			1        &v_1^{(1)}&w_1^{(1)}&\cdot&\cdot&\cdot\\
			\cdot    &v_m^{(1)}&w_m^{(1)}&1    &\cdot&\cdot\\
			\cdot    &\cdot    &1    &v_1^{(2)}&w_1^{(2)}&\cdot\\
			\cdot    &\cdot    &\cdot&v_m^{(2)}&w_m^{(2)}&1    \\
			\cdot    &\cdot    &\cdot&\cdot&1    &v_1^{(3)}\\
			\end{array}}
		\right)
		$
	\end{center}
	By solving the following equations
	\begin{equation}\label{eq:blockAYbar}
	\tilde{A}\overline{Y} = VP,
	\end{equation}
	\begin{equation}\label{eq:Zppt}
		\overline{Z}y = E^T\tilde{x},
	\end{equation}
	we finally obtain the solution $x$
	\begin{equation}\label{eq:xppt}
	x = \tilde{x} - \overline{Y}y.
	\end{equation}

	For a strictly diagonal dominant TDMA whose diagonal elements at the $j$-th row satisfy
\begin{center}
	$\abs{b_j}>\abs{a_j}+\abs{c_j}$,
\end{center}
the off-diagonal elements of the $Z$ matrix, $v_m^{(i)}$ and $w_1^{(i)}$ converge to zero when $n\gg p$. Then the matrix $Z$ can be approximated as a block-diagonal matrix with $2\times2$ blocks which can be solved without communication. Thus, the PDD method solves Eqs.~(\ref{eq:blockAx}), (\ref{eq:blockAY}), (\ref{eq:Zpdd}) and (\ref{eq:xpdd}) locally in each GPU with a small amount of neighbor-to-neighbor communication. As will be shown later, this method has an excellent scalability thanks to the small communication cost.

	On the other hand, the PPT method makes no approximation, so it can be applied to general tridiagonal systems. In this method, Eqs.~(\ref{eq:blockAx}), (\ref{eq:blockAYbar}) and (\ref{eq:xppt}) are solved locally on independent GPUs, but the same Eq.~(\ref{eq:Zppt}) needs to be solved by every GPU. Thus an all-gather communication is required for creating the $\overline{Z}$ matrix on each GPU.\\

	\textit{PDD method for the momentum equation}

	For the present study, PDD method is used in solving the momentum equation along the wall-normal $y$-direction. This is possible because tridiagonal matrices resulting from the ADI method have a strictly diagonal dominant property such that
	\begin{center}
		$\abs{b_j}=\abs{a_j}+\abs{c_j}+1$.
	\end{center}
	This ensures that the solution of the momentum equation from the PDD method matches the exact solution within machine accuracy.
	
	Note that fine-grained parallelism is essential when using this method on GPUs. The PDD method establishes a scalable domain decomposition at the coarse level, but its performance depends on how the tridiagonal systems of Eqs.~(\ref{eq:blockAx}) and (\ref{eq:blockAY}) are solved. To do so, we utilize the 4-level parallelism used in \cite{ha2018gpu} and extend this up to 5 levels by batching Eq.~(\ref{eq:blockAY}). A hybrid Cyclic Reduction (CR) + Parallel Cyclic Reduction (PCR) algorithm \cite{zhang} is used which is provided in the cuSPARSE library as \texttt{cusparseDgtsv\_nopivot} \cite{cusparse}. Details are described in Algorithm~\ref{alg:PDD}.\\
	
	\textit{PPT method for the Poisson equation}

	For the present study, PPT method is used in solving $y$-directional TDMAs of the Poisson equation. Its major diagonal, $\boldsymbol{b}$ is made of off-diagonals, $\boldsymbol{a}+\boldsymbol{c}$ plus the modified wavenumbers coming from the half-cosine transform in the $x$-direction followed by the Fourier transform in the $z$-direction. Thus the matrices may have only a slight diagonal dominance of $\abs{b_j} = \abs{a_j}+\abs{c_j}+\epsilon$ with a small $\epsilon$ depending on the size of modified wavenumbers. The authors have found that $\epsilon$ may easily fall down to $O(10^{-6}) \sim O(10^{-10})$ for which the PDD method has given inaccurate results.

	Similar to the PDD method, it is important to use fine-grained parallelism when solving Eqs.~(\ref{eq:blockAx}) and (\ref{eq:blockAYbar}). Methods used to solve the Poisson equation in \cite{ha2018gpu} are employed in which a parallel tridiagonal solver with diagonal pivoting is used \cite{chang2012}. \texttt{MPI\_ALLGATHER} is used to collect data for configuring the $\overline{Z}$ matrix in each GPU.

	\section{Performance results}
	Numerical experiments are conducted to evaluate the scalability of the present multi-GPU solver. The GPU code runs on an IBM Power System S822LC for High Performance Computing. This server has two octa-core Power8 CPUs and four Tesla P100 GPUs with NVLink interconnect. The code is compiled with an \texttt{-O2} optimization of the PGI Fortran Compiler version 18.4. Performance is tested in simulations of a flat-plate boundary layer whose boundary conditions are given in section 2.	

\begin{landscape} 
		\begin{figure}
	\centering
	\includegraphics[width=200mm]{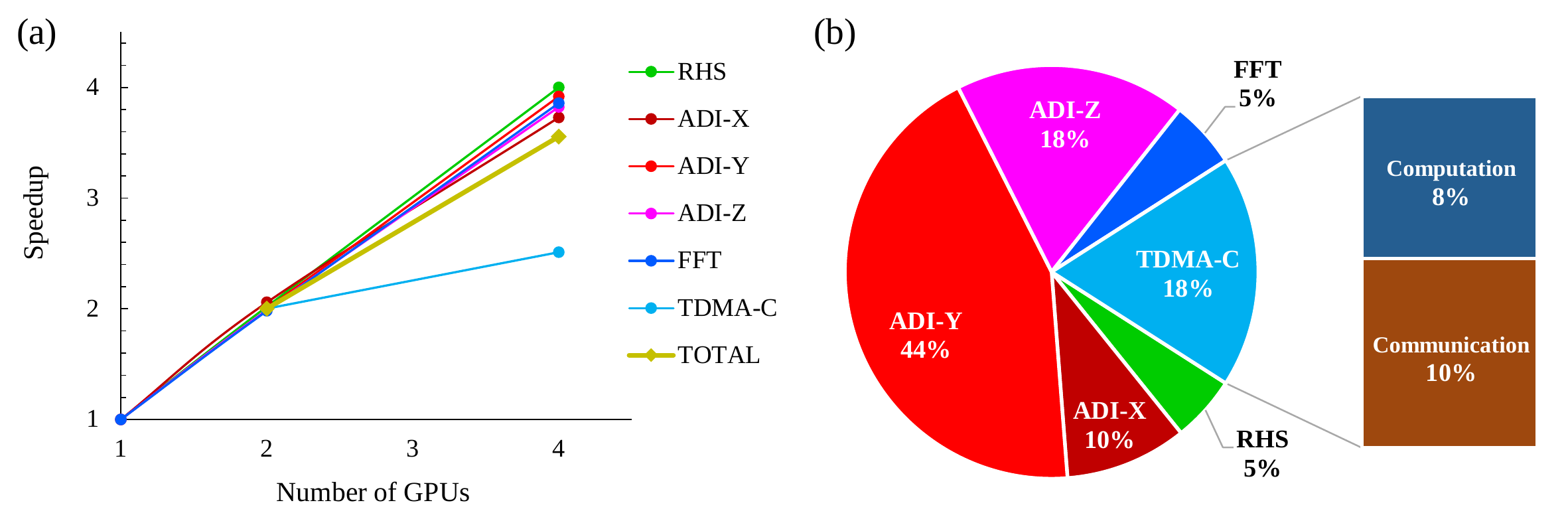}
	\caption{Multi-GPU performance of the present code. (a) Scaling of each component of the Navier-Stokes equations. Speedup results have been measured on 135M grid points. The gold $\blacklozenge$ marker shows the scaling of the entire code. (b) Relative importance of each component based on the wall-clock time. Note that TDMA-C which shows the worst scaling takes up 18\% of the entire solver, and it spends more than half of its time on all-gather communication.}
	\label{fig:scalability}
		\end{figure}
	\end{landscape}

	Scaling of the main components of the semi-implicit fractional-step method is shown for 4 GPUs in Fig.~\ref{fig:scalability}. Speedup has been measured on $4096\times256\times128 = 134$M cells. In the $y$-directional domain decomposition, the right-hand side momentum equation (RHS), the ADI method in the $x$-direction (ADI-X) and the $z$-direction (ADI-Z), and FFT in $x$ and $z$ directions are computed independently on each GPU without communication. Thus, strong scaling has been obtained as expected. The more interesting part is the performance of the ADI method in the $y$-direction (ADI-Y), and inversion of the complex-numbered TDMA of the Poisson equation (TDMA-C), for which PDD and PPT methods are applied, respectively. Thanks to the small communication cost of the PDD method, ADI-Y scales very well on multiple GPUs. Given that the ADI method is the main bottleneck of the present fractional-step method, the PDD method has drastically increased the overall scalability of the solver. On the contrary, TDMAs of the Poisson equation have weak scaling properties. This is attributed to the all-gather communication of the PPT method which is shown to take up more than half of the total time taken to invert the TDMAs (Fig.~\ref{fig:scalability}(b)). However note that this communication cost represents $10\%$ of the total time, which is much less than the cost required for global transpose that usually amounts to $30\%\sim40\%$. As a result, an efficiency of 0.89 is achieved for the entire solver as shown in the golden curve of Fig.~\ref{fig:scalability}.
	
\begin{figure}
	\centering
	\includegraphics[width=140mm]{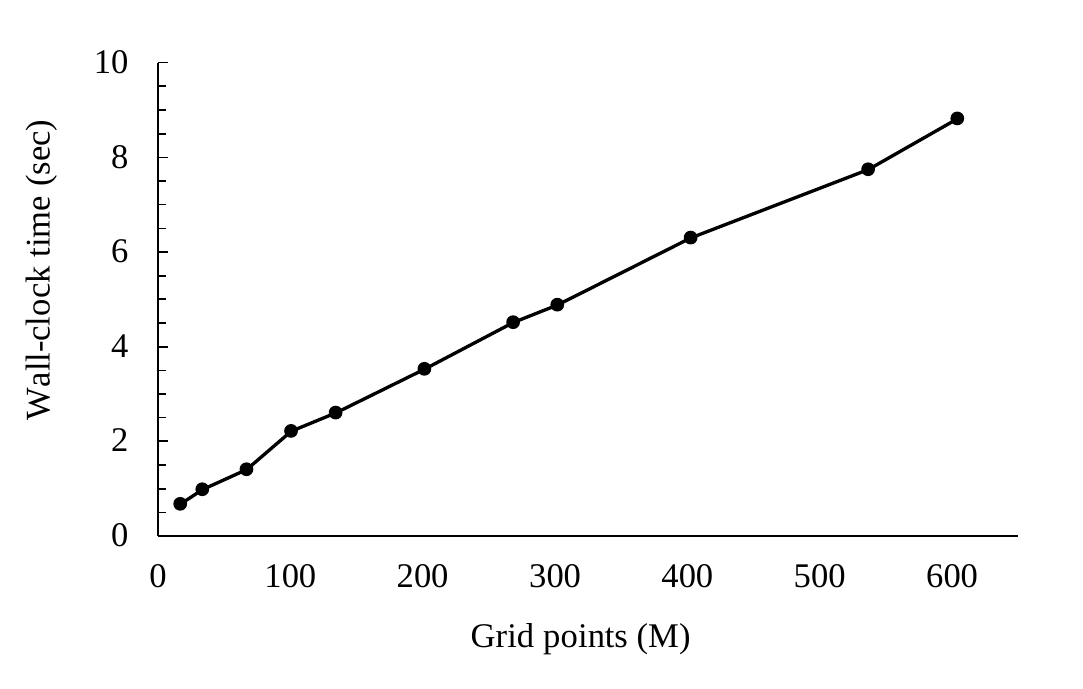}
	\caption{Wall-clock time of a time-step on various grid sizes using four GPUs. Specific values given in Table~\ref{tab:wallclock}.}
	\label{fig:wallclock}
\end{figure}

\begin{landscape} 
	\begin{figure}
		\centering
		\includegraphics[width=200mm]{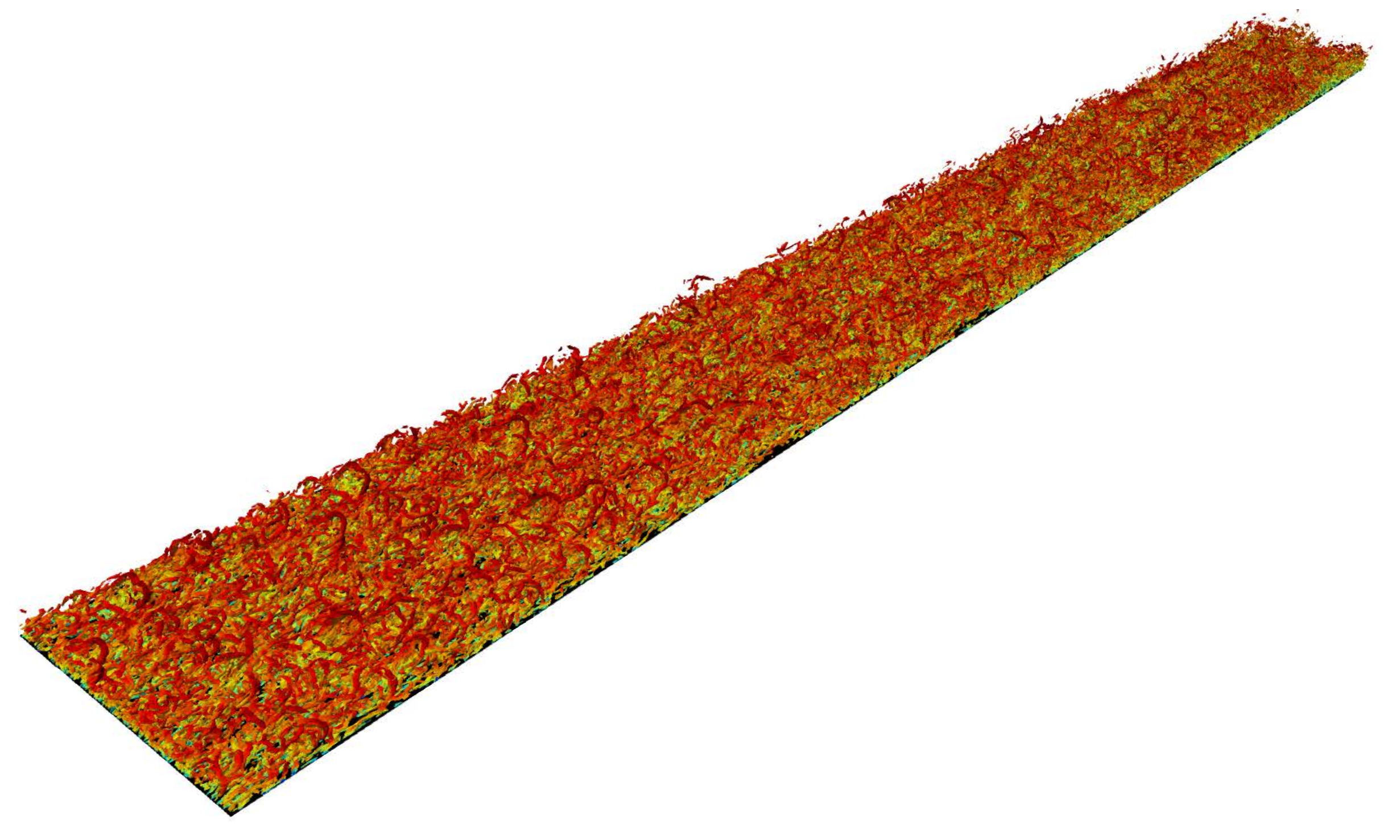}
		\caption{Turbulent boundary layer over a flat plate at inlet $Re_{\theta}=1000$. 607M grid points have been computed using four P100 GPUs. Q-criterion is used for visualization.}
		\label{fig:zpg1000}
	\end{figure}
\end{landscape}

 	Performance on different grid sizes is investigated by measuring average wall-clock time for one time-step using 4 GPUs. Collected data are listed in Table \ref{tab:wallclock} and plotted in Fig.~\ref{fig:wallclock}. A fairly linear increase of computation time is shown as the grid size is increased, which implies that communication cost does not increase significantly as the problem size increases. Note that a sudden increase in the slope of the curve occurs whenever the grid cell size contains a multiple of 3. A similar phenomenon has previously been observed in the single-GPU code. This is because the solver spends most of its time on reduction algorithms, which are known to perform best when the problem size is a power of 2 \cite{ha2018gpu}. Using the largest grid tested (607M), a turbulent flat-plate boundary layer at $Re_{\theta}=1000$ has been simulated (Fig.~\ref{fig:zpg1000}). For a fixed CFL=1.0, the average time-step size was 0.022, and it took roughly 2 days to advance a flow-through time. 

	\section{Conclusion}
	A multi-GPU solver using the semi-implicit fractional-step method is developed for DNS of wall-bounded incompressible flow. Global transpose required for extending the ADI and Fourier-transform based direct methods to multiple GPUs is found to be impractical. A one-dimensional domain decomposition in the wall-normal $y$-direction is proposed, which allows us to compute FFT and ADI method in $x$ and $z$ directions locally on each GPU without communication. Systems of $y$-directional TDMAs distributed across multiple GPUs are solved by implementing PDD and PPT methods in a GPU-friendly way. An algorithm for maximizing GPU workload is provided, which combines the coarse-grained parallelism of the PDD method and the fine-grained parallelism of individual TDMAs. The momentum equation with the PDD method shows a strong scaling while the Poisson equation with the PPT method shows a weak one. An overall efficiency of 0.89 is obtained for 4 GPUs. A turbulent flat-plate boundary layer has been simulated on 607M grid points using only $4\times$P100 GPUs of a single node, which shows a promising potential for large-scale DNS on GPU clusters.
	
	
	\section*{Acknowlegements}
	This research was supported by the Samsung Research Funding Center of Samsung Electronics (SRFC-TB1703-01) and National Research Foundation of Korea grant funded by the Korea government (NRF-2017R1E1A1A03070514).
	
	
	

\pagebreak
	\bibliographystyle{elsarticle-harv} 
	\bibliography{bibfile}
	
	

	\begin{algorithm}[H]
	{\tiny
	\label{alg:PDD}
	\caption{The $y$-directional ADI using PDD method}
	
	\SetKwFunction{CalcKappa}{calcKappa}
	\SetKwFunction{ConfDiag}{configureV}
	\SetKwFunction{Allocate}{allocate}
	\SetKwFunction{Deallocate}{deallocate}
	\SetKwFunction{cusparse}{cusparseDgtsv\_nopivot}
	\SetKwFunction{Sync}{cudaDeviceSynchronize}
	\SetKwInOut{Diag}{LHS diagonals}
	\SetKwInOut{RHS}{RHS diagonals}
	\SetKwInOut{GridSize}{Grid cell size}
	\SetKwInOut{Input}{Batch size}
	\SetKwInOut{Temp}{temp}
	\SetKwInOut{Output}{Final solution}
	
	\BlankLine
	\Input{$\kappa$ from \cite{ha2018gpu}}
	\Diag{$a$, $b$, $c$}
	\RHS{$d_1$, $d_2$, $d_3$ \hspace{4pt} ! from each $u,v,w$ momentum equations}
	\GridSize{$N_x, N_y, N_z$}
	
	\BlankLine
	$r=$~\texttt{rank}\\
	$m=N_y/p$\\
	$M=m*N_x*\kappa$\\
	\Allocate{$R(9M)$}\\
	$R(1:M) = d_1$\\
	$R(1+3M:4M) = d_2$\\
	$R(1+6M:7M) = d_3$\\
	$kfin = N_z / \kappa$\\
	\For{$1$ \KwTo $kfin$}{
		\textcolor{ForestGreen}{\textbf{
	! Step 1. configure the matrix $V$ of Eq.~(\ref{eq:blockAY}) in $R$.}}\\
		$R(1+M:2M)$, $R(1+4M:5M)$, $R(1+7M:8M)$ $\leftarrow$ \texttt{configureAcorners($a$)}\\
		$R(1+2M:3M)$, $R(1+5M:6M)$, $R(1+8M:9M)$ $\leftarrow$ \texttt{configureCcorners($c$)}\\
		\textcolor{ForestGreen}{\textbf{
	! Step 2. set MPI boundaries to zero.}}\\
		\ForEach{$i=1:N_x$ and $k=1:\kappa$}
		{
			$a(j=1,i,k) = 0.$; \hspace{2pt} $c(j=m,i,k) = 0.$\\
		}
		\textcolor{ForestGreen}{\textbf{
	! Step 3. solve for $\tilde{x},$~$Y$ of Eqs.~(\ref{eq:blockAx}),~(\ref{eq:blockAY}).}}\\
		\texttt{call} \cusparse{\texttt{handle}, $mN_x$, $3\kappa$, $a,b,c$, $R(1:3M)$, $mN_x$}\\
		\texttt{call} \cusparse{\texttt{handle}, $mN_x$, $3\kappa$, $a,b,c$, $R(1+3M:6M)$, $mN_x$}\\
		\texttt{call} \cusparse{\texttt{handle}, $mN_x$, $3\kappa$, $a,b,c$, $R(1+6M:9M)$, $mN_x$}\\
		\textcolor{ForestGreen}{\textbf{
	! Step 4. send $\tilde{x}_1^{(r)}$, $v_1^{(r)}$ of the $r$-th GPU to the $(r-1)$-th GPU.}}\\
		\ForEach{$i=1:N_x$ and $k=1:\kappa$ and $u=1:3$}
		{
			\texttt{sbuf\_x1}(i,k,u) $\leftarrow$ pack elements of $R(1:M), R(1+3M:4M), R(1+6M:7M)$ at $(j=1,i,k)$\\
			\texttt{sbuf\_v1}(i,k,u) $\leftarrow$ pack elements of $R(1+M:2M), R(1+4M:5M), R(1+7M:8M)$ at $(j=1,i,k)$\\
		}
		\texttt{cudaStreamSynchronize}\\
		\texttt{call} \texttt{MPI\_SENDRECV}~(\texttt{sbuf\_x1}, $3N_x\kappa$, $\cdots$)\\
		\texttt{call} \texttt{MPI\_SENDRECV}~(\texttt{sbuf\_v1}, $3N_x\kappa$, $\cdots$)\\
		\textcolor{ForestGreen}{\textbf{
	! Step 5. compute $y$ except for the last GPU ($r=p-1$).}}\\
		\If{$r\neq (p-1)$}{
		Compute $y=(y_{2r+1}, y_{2r+2})^T$ of Eq.~(\ref{eq:Zpdd}) using the formula for the inverse of a $2\times2$ matrix.}
		\textcolor{ForestGreen}{\textbf{
	! Step 6. send $y_{2r-1}$ from the $r$-th GPU to the $(r+1)$-th GPU.}}\\
		Similar to step 4 above.\\
		\textcolor{ForestGreen}{\textbf{
	! Step 7. compute Eq.~(\ref{eq:xpdd})}}\\
		$Yy = \left({\begin{array}{cc}
			\boldsymbol{v}^{(r)} & \boldsymbol{w}^{(r)}
			\end{array}} \right)
		\left(	{\begin{array}{c}
			y_{2r-1}\\
			y_{2(r+1)}\\
			\end{array}} \right)$, where $y_{-1}=0$, $y_{2p}=0$
	}
	\BlankLine
	$d_1 = R(1:M)$\\
	$d_2 = R(1+3M:4M)$\\
	$d_3 = R(1+6M:7M)$\\
	\Deallocate{$R$}\\
	\Output{$d_1$, $d_2$, $d_3$}
	}
	\end{algorithm}

	\newpage
	\listoftables
	\listoffigures
	\newpage

	\newpage
	{\renewcommand{\arraystretch}{1.3}
	\begin{table}
		\centering
		\begin{tabular*}{1.0\linewidth}{@{\extracolsep{\fill}}lll}
			\toprule
			Grid cell dimension & Total grid points (M) & Wall-clock time (sec) \\
			\midrule
			 $256\times256\times 256$& 16  &  0.67\\
			 $512\times256\times 256$& 33  &  0.98\\
			 $512\times256\times 512$& 67  &  1.40\\
			 $768\times256\times 512$& 101 &  2.21\\
			$1024\times256\times 512$& 135 &  2.60\\
			$1536\times256\times 512$& 202 &  3.52\\
			$1024\times256\times1024$& 270 &  4.51\\
			$1536\times384\times 512$& 303 &  4.88\\
			$1536\times256\times1024$& 404 &  6.30\\
			$2048\times256\times1024$& 539 &  7.74\\
			$3072\times256\times 768$& 607 &  8.82\\
			\bottomrule
		\end{tabular*}
		\caption{Wall-clock time(sec) measured using four Tesla P100 GPUs. For each grid size, average computation time is measured for one time-step (three sub-steps). Grid dimension is given as the number of cells in each $x$, $y$ and $z$ direction. Total number of grid points are written in millions.}
		\label{tab:wallclock}
	\end{table}}
	\clearpage

\pagebreak

	\clearpage

	\clearpage

%
		
\end{document}